\documentclass[aps,prc,fleqn,floatfix,twocolumn,twoside]{revtex4}

\usepackage[dvipdfm]{graphicx}

\usepackage{amsmath,amssymb,amsfonts}

\usepackage{color}

\begin{document}

\title{Explanation of Di-jet asymmetry in Pb+Pb collisions at the Large Hadron Collider}

\author{Guang-You Qin and Berndt M\"uller}
\affiliation{Department of Physics, Duke University, Durham, NC 27708, USA}

\date{\today}
\begin{abstract}

We investigate the medium modification of a partonic jet shower traversing in a hot quark-gluon plasma.
We derive and solve a differential equation that governs the evolution of the radiated gluon distribution as the jet propagates through the medium.
Energy contained inside the jet cone is lost by dissipation through elastic collisions with the medium and by scattering of shower partons to larger angles.
We find that the jet energy loss at early times is dominated by medium effects on the vacuum radiation, and by medium-induced radiation effects at late times.
We compare our numerical results for the nuclear modification of the di-jet asymmetry with that recently reported by the ATLAS Collaboration.

\end{abstract}
\maketitle

Quarks and gluons scattered to large transverse momentum $p_T$ during the early stage of a relativistic heavy-ion collision are regarded as powerful probes of the transient quark-gluon plasma produced in these reactions \cite{Majumder:2010qh}.
Such hard partons interact with the traversed matter and lose some of the initial energy by elastic and inelastic collisions with constituents of the medium \cite{Bjorken:1982tu, Zakharov:1996fv, Baier:1996kr}.
The essential aspects of parton energy loss have been confirmed by observations of strong high-$p_T$ single hadron suppression in nucleus-nucleus collisions at the Relativistic Heavy-Ion Collider (RHIC) \cite{Adcox:2001jp, Adler:2002xw}.
Other observations supporting this interpretation include the suppression of back-to-back di-hadron and correlated back-to-back photon-hadron emission \cite{Adler:2002tq, Adare:2009vd, Abelev:2009gu}.
These phenomena are commonly referred to as ``jet quenching''.
Although direct calorimetric measurement of jets has been performed at RHIC,
detailed studies of the medium modification of jet showers are constrained by limited center-of-mass (c.m.) energy \cite{Bruna:2009em, Lai:2009zq, Vitev:2009rd}.

The launch of the heavy-ion program at the CERN Large Hadron Collider (LHC) has extended the available c.m. energy range by more than a factor of ten, enabling the investigation of medium effects on jets with transverse energies $E_T$ in excess of 100~GeV.
Very recently, the ATLAS Collaboration reported first results on the modification of di-jet properties in Pb+Pb collisions at $\sqrt{s_{NN}}=2.76$~TeV \cite{Collaboration:2010bu}.
The most spectacular aspect of these results is the occurrence of extreme energy imbalances between two correlated jets emitted in opposite azimuthal directions around the beam axis.
Similar preliminary results were reported by the CMS Collaboration \cite{Chatrchyan:2011sx}.
These results provide a challenge for theoretical descriptions of the parton shower evolution in hot QCD matter.
While a qualitative analysis of the physics processes responsible for the observed large di-jet asymmetry was presented in Ref. \cite{CasalderreySolana:2010eh}, no quantitative description of the experimental results has been reported to date.

In this work, we study the medium modification of a parton shower in an expanding quark-gluon plasma and calculate the energy loss experienced by the shower within a cone angle defined as $R=\sqrt{(\Delta \phi)^2 + (\Delta \eta)^2}$.
The total loss of energy from the jet cone is the sum of energy loss incurred by the leading parton and its radiated gluons, as well as the energy carried by gluons that are scattered out of the jet cone.
The leading parton may transfer energy to the medium by collisions with matter constituents; this energy loss is quantified by the transport coefficient $\hat{e} = dE/dt$, denoting the collisional energy loss per unit time.
The leading parton also radiates gluons due to scattering in the medium; this effect contributes to the depletion of energy contained within the jet cone if the radiated gluons are emitted outside the cone.
In addition, the radiated gluons (from both vacuum and medium-induced radiation) originally contained inside the jet cone not only suffer collisional energy loss, but they may be kicked out of the jet cone by multiple scatterings in the medium after their emission.
The transverse momentum broadening of the jet shower is characterized by the transport coefficient $\hat{q} = d{(\Delta p_\perp)^2}/dt$ \cite{Baier:1996kr, Majumder:2007hx}.
Here we relate these two transport coefficients by $\hat{q} = 4T \hat{e}$, assuming a medium close to local thermal equilibrium and the applicability of the fluctuation-dissipation theorem \cite{Qin:2009gw}.

We now derive a differential equation that describes how the radiated gluon distribution $f_g(\omega, k_\perp^2,t) = dN_g(\omega, k_\perp^2, t)/d\omega dk_\perp^2$ changes when the jet shower propagates through the medium.
Here $\omega$ denotes the gluon energy and $k_\perp$ the transverse momentum with respect to the jet axis.
The resulting modified gluon distribution is used to calculate the energy loss from radiated gluons inside the jet cone.
The total energy loss from the jet cone is the sum of the energy loss from the leading parton and the radiated gluons, $\Delta E = \Delta E_{\rm L} + \Delta E_{g}$.

Consider a partonic shower that is propagating through a hot quark-gluon plasma.
During their propagation, the radiated gluons will transfer energy into the medium by elastic collisions and accumulate transverse momentum in the process.
The interaction between the leading parton and the medium may induce additional gluon radiation given by the rate $dN_g^{\rm med}/d\omega dk_\perp^2 dt$.
After their emission, these gluons may also lose energy and accumulate transverse momentum.
Combining these three contributions, we obtain the following differential equation for the radiated double differential gluon distribution:
\begin{equation}
\label{eq:dG/dt}
\frac{d}{dt}f_g(\omega, k_\perp^2, t) = \hat{e} \frac{\partial f_g}{\partial \omega}
  + \frac{1}{4} \hat{q} {\nabla_{k_\perp}^2 f_g} +  \frac{dN_g^{\rm med}}{d\omega dk_\perp^2 dt} .
\end{equation}
Note that the $\omega$ dependence of the jet shower evolution has been discussed in Ref. \cite{Neufeld:2009ep}, where the medium response to the energy deposition of a jet shower was studied (see also Ref. \cite{Qin:2009uh}).

The above equation must be supplemented with an equation for the initial gluon distribution, $f_g(\omega, k_\perp^2,t_i)$ generated by vacuum showering of the primary parton before interacting with the thermal medium at time $t_i$.
The vacuum radiation spectrum is generated from PYTHIA \cite{Sjostrand:2007gs}, and the gluon distribution $f_g^{\rm vac}(\omega, k_\perp^2, t_i)$ at the initial time is obtained by requiring that only gluons with a formation time $\tau_f = 2Ex(1-x)/k_\perp^2$ smaller than $t_i$ are radiated, where $x=\omega/E$ denotes the energy fraction of the radiated gluons.

In the higher-twist formalism of jet quenching, the rate for medium-induced gluon radiation is obtained as \cite{Wang:2001ifa, Majumder:2009ge}:
\begin{eqnarray}
\label{eq:Gmed}
\frac{dN_g^{\rm med}}{d\omega dk_\perp^2 dt} = \frac{2\alpha_s}{\pi} \frac{x P(x) \hat{q}(t)}{\omega k_\perp^4}   \sin^2 \frac{t - t_i}{2\tau_f} ,
\end{eqnarray}
where $P(x)$ is the vacuum splitting function.
The transport coefficient $\hat{q}$ depends on time because of the expansion of the medium as well as the changing position of the primary parton.
When solving for the radiated gluon distribution $f_g(\omega, k_\perp^2, t)$, we impose a lower cut-off $\omega_{\rm min} = 2$~GeV on the radiated gluon energy,
and any radiated gluon whose energy drops below $\omega_{\rm min}$ is considered as part of the medium.
This energy cut-off is about four times the highest temperature of the medium created in Pb+Pb collisions at the LHC.

As we already discussed, our initial condition for the radiated gluon distribution is taken as the vacuum radiation shower obtained at time $t_i$, after which we start the medium evolution of the gluon distribution.
This is motivated by the observation that the vacuum contribution to the primary parton virtuality dominates at earlier time while the virtuality contribution from the medium takes over at later times \cite{Muller:2010pm}.
Furthermore, interference effects among subsequent gluon emissions ensure that the radiated gluons in the vacuum are angular ordered, {\em i.e.}\ that gluons emitted at late times are radiated into small angles within the jet cone.
Vacuum radiation at late times will thus not be sufficiently modified by the medium to contribute to the energy loss of the shower outside the jet cone.
We note, however, that the vacuum radiation would need to be included as a source term in Eq.~(\ref{eq:dG/dt}) in order to account for the interference between vacuum and medium-induced radiation.

After solving Eq. (\ref{eq:dG/dt}) and obtaining the time evolution of radiated gluon distribution $f_g(\omega, k_\perp^2, t)$, we may calculate the final energy of the gluons contained inside the jet cone:
\begin{eqnarray}
\label{eq:Eg}
E_g(t_f, R) = \int_R \omega d\omega dk_\perp^2 f_g(\omega, k_\perp^2, t_f)  , \ \ \
\end{eqnarray}
where the subscript $R$ indicates that the integration is performed over the interior of the jet cone with the condition $k_\perp/\omega \leq R$.
The total energy inside the jet cone is the sum of the energies of the leading parton and the gluons inside the cone:
\begin{eqnarray}
\label{Ejet}
E_{\rm jet}(t_f, R) = E_{\rm L}(t_f) + E_g(t_f, R) ,
\end{eqnarray}
where the final energy for the leading parton is given by
\begin{eqnarray}
\label{eq:EL}
E_{\rm L}(t_f) &=& E_{\rm L}(t_i) - \int_{t_i}^{t_f} \hat{e}(t) dt
\nonumber \\
&& - \int \omega d\omega dk_\perp^2 dt\, \frac{dN_g^{\rm med}}{d\omega dk_\perp^2 dt}  .
\end{eqnarray}
The above equation describes the time evolution of leading parton energy in the medium.
The momentum broadening of the leading parton in the medium may be included in a similar way.
The leading parton energy at the starting time $t_i$ is obtained as
\begin{eqnarray}
E_{\rm L}(t_i) = E_{\rm jet}(0) - \int \omega d\omega dk_\perp^2 f_g(\omega, k_\perp^2, t_i).
\end{eqnarray}
The complete reduction of energy contained within the jet cone is
\begin{eqnarray}
\label{eq:DeltaE}
\Delta E = E_{\rm jet}(t_i, R) - E_{\rm jet}(t_f, R) .
\end{eqnarray}
The initial energy within the jet cone is the sum of leading parton energy and the energy of the gluons inside the cone:
\begin{equation}
\label{eq:Ejet-2}
E_{\rm jet}(t_i, R) = E_{\rm L}(t_i) + \int_R \omega d\omega d k_\perp^2 f_g(\omega, k_\perp^2, t_i) .
\end{equation}
The final expression for the energy lost from the jet cone is:
\begin{eqnarray}
\Delta E &=& \int_{R} \omega d\omega dk_\perp^2  \left[f_g(\omega, k_\perp^2, t_i)  - f_g(\omega, k_\perp^2, t_f) \right]
\nonumber\\
\label{eq:DeltaE-2}
&& + \int \omega d\omega dk_\perp^2  dt\, \frac{dN_g^{\rm med}}{d\omega dk_\perp^2 dt}
+ \int_{t_i}^{t_f} \hat{e}(t) dt .
\end{eqnarray}
The first three terms describe the energy loss due to radiated gluons, while the last term is the collisional energy loss by the leading parton.
Equation (\ref{eq:DeltaE-2}) is quite general and can be understood if one splits the medium-induced radiation term into two parts: the radiated energy inside and outside the jet cone.
Then the first and third term represent the gluon energy gains inside the cone, the second term is the remaining gluon energy inside the cone, and the last two terms represent the energy loss from the leading parton.

\begin{figure}[htb]
\includegraphics[width=1.\linewidth]{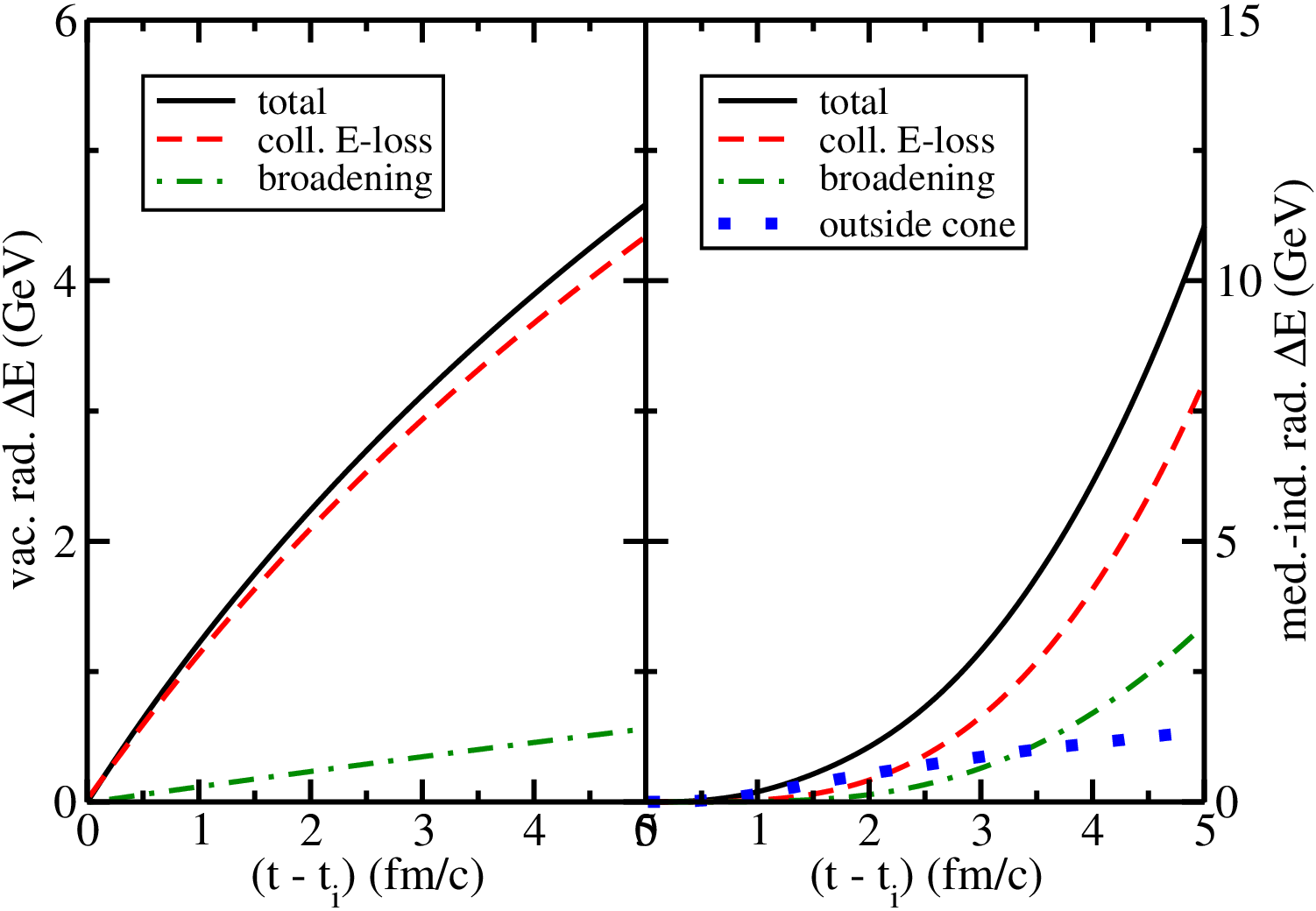}
\caption{(Color online)
The energy loss experienced by a 100~GeV quark jet traversing a brick of quark-gluon plasma with constant temperature $T=250$~MeV.
Contributions due to different mechanisms of energy loss from vacuum radiation (left) and medium-induced radiation (right) are compared.
}
\label{brick_vac_med}
\end{figure}

We first apply the above formalism to the energy loss of a jet in a static medium.
The result is shown in Fig.~\ref{brick_vac_med}, where we show the contributions from vacuum radiated gluons (left) and medium-induced radiated gluons (right) to the total energy loss of the jet defined by a cone angle $R=0.4$.
We begin with a quark jet with initial energy $E=100$~GeV which radiates gluons in vacuum before $t_i=1$~fm/c, when we turn on the medium modification for the vacuum shower.
The figure shows the additional energy loss experienced by the jet shower inside the medium.
Here we set the temperature of the medium to be $T=250$~MeV and the transport coefficient to be $\hat{q}=0.7$~GeV$^2$/fm.
The figure shows that vacuum radiation dominates the energy loss at early times; medium-induced radiation later takes over due to its much stronger length dependence.
For both vacuum and medium-induced radiation, we observe that the most significant contribution originates from collisional energy loss experienced by radiated gluons.
For medium-induced radiation, the transverse momentum broadening of shower gluons presents another sizable contribution, while the contribution from radiation outside the jet cone is small.

\begin{figure}[htb]
\includegraphics[width=1.\linewidth]{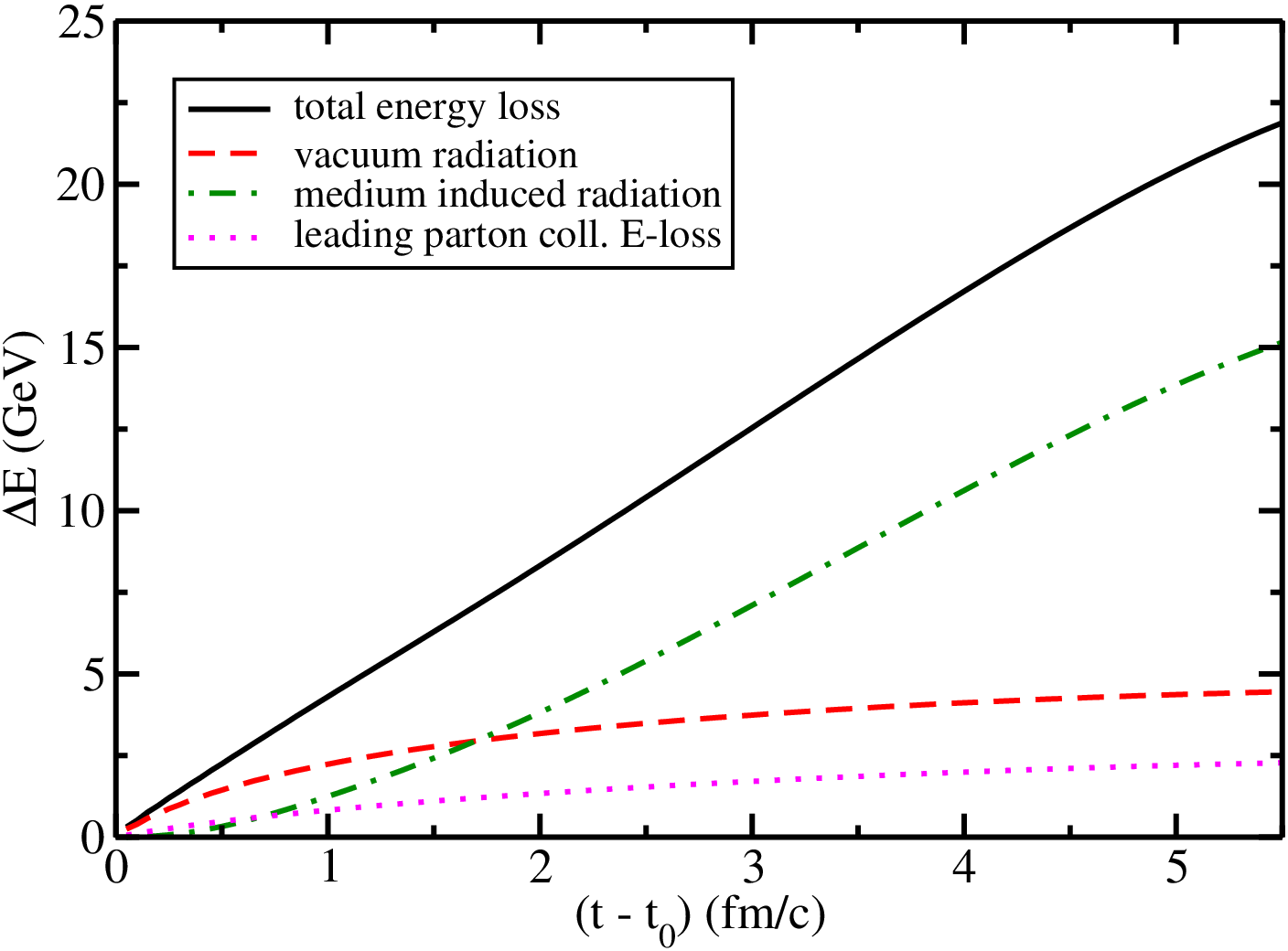}
\caption{(Color online)
The energy loss experienced by a $100$~GeV quark jet initially produced in the center of a head-on Pb+Pb collisions at $\sqrt{s_{NN}}=2.76$~TeV at the LHC.
Contributions from various components are shown for comparison: total (solid line), vacuum radiation (dashed line), medium-induced radiation (dot-dashed line) and leading parton collisional energy loss (dotted line).
}
\label{lhc_bj_dE}
\end{figure}

We now apply our formalism to calculate the jet cone energy loss and the medium modification of the di-jet asymmetry in Pb+Pb collisions at the LHC reported by the ATLAS Collaboration.
To proceed, we scale the transport coefficients according to the temperature of the medium, $\hat{q} \propto T^3$, with a constant factor adjusted to fit the experimental data.
We model the space-time profile of the temperature $T(\vec{r}_\perp, t)$ at midrapidity as follows.
The initial spatial profile for the entropy density $s\sim T^3$ of the medium is set to be proportional to the density of participating nucleons in the colliding nuclei.
The nuclear density distributions are taken as Woods-Saxon profiles.
The medium created in Pb+Pb collisions at $\sqrt{s_{NN}}=2.76$~TeV is assumed to thermalize at $t_0 = 0.6$~fm/$c$, at which time the temperature of the hottest point in central collisions is set to $T_0 = 520$~MeV, a factor of $1.3$ larger than the highest temperature in Au-Au collisions at $\sqrt{s_{NN}}=200$~MeV at RHIC.
This assumption is consistent with the measurement of the total multiplicity in Pb+Pb collisions at the LHC, which is a factor of $2.2$ larger than that measured in Au+Au collisions at RHIC \cite{Aamodt:2010pb}.
The time evolution of medium is modeled by a one-dimensional boost-invariant expansion, {\em i.e.}, the temperature falls with time as $t^{-1/3}$.
The shower evolution is terminated when the medium temperature drops to $T_c=160$~MeV, below which we assume the partonic shower suffers no energy loss in the medium.

In Fig.~\ref{lhc_bj_dE} we show the medium-induced energy loss experienced by a quark jet with initial energy $E=100$~GeV produced in the center of the medium created in a head-on collision of two Pb nuclei at $\sqrt{s_{NN}}=2.76$~GeV.
We observe that, due to falling temperature and the varying transport coefficient $\hat{q}$ along the jet path, the length dependence of the energy loss become much weaker than for a static medium as shown in Fig.~\ref{brick_vac_med}.
Such a jet, which is produced at the center of the collision region, experiences vacuum radiation before $t_i=t_0=0.6$~fm/c
and then interacts with the medium for $t_f- t_0 \approx 5.5$~fm/c before the medium temperature drops below $T_c=160$~MeV.
The total energy loss from the jet cone during this period is $\Delta E \approx 22$~GeV.
Before $t-t_0 \approx 1.7$~fm/$c$, the energy loss is dominated by vacuum radiated gluons produced before the medium evolution starts at $t_0$.
After this time, medium-induced radiation takes over and finally contributes about $70\%$ of the total energy loss.

\begin{figure}[htb]
\includegraphics[width=1.\linewidth]{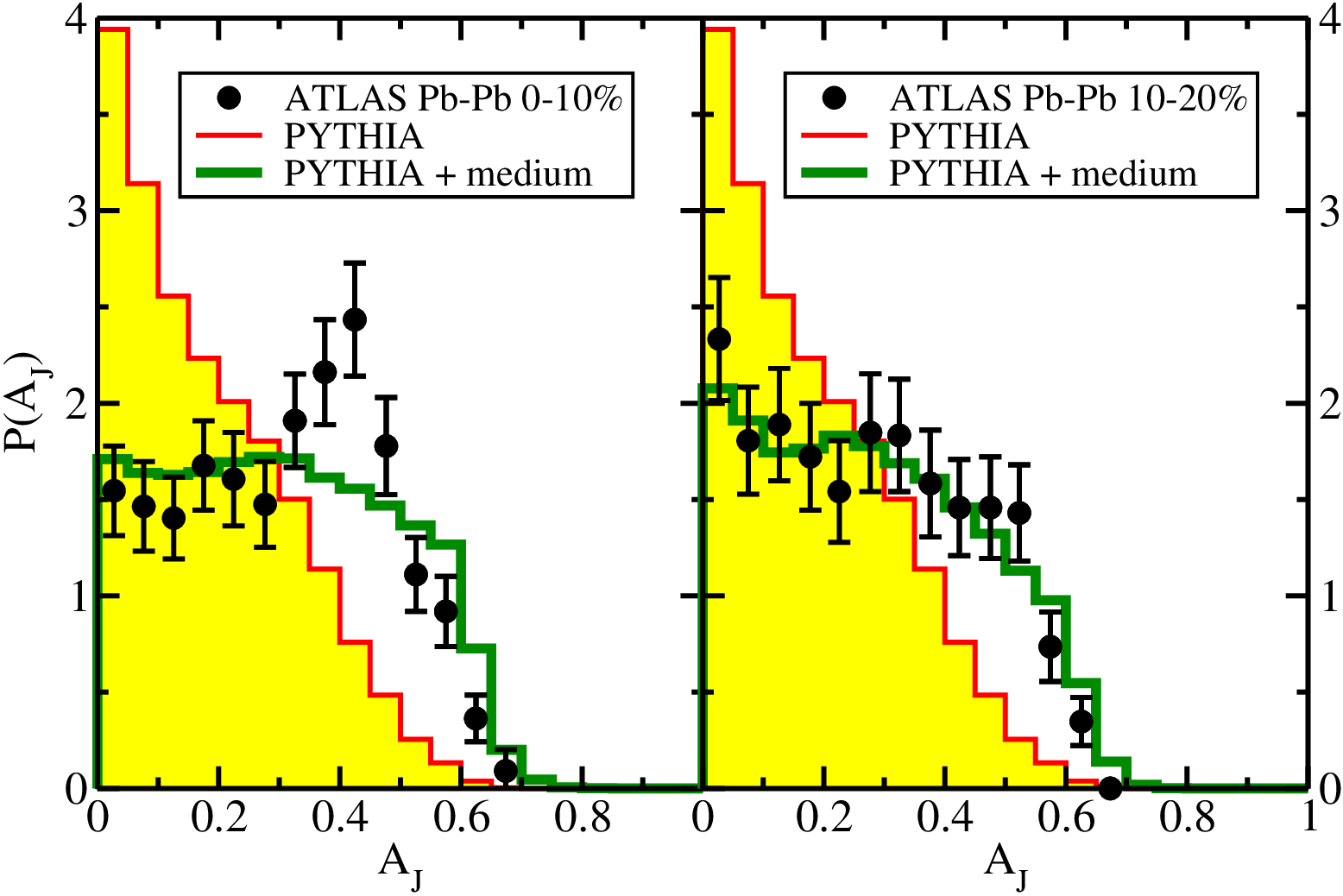}
\caption{(Color online)
Distribution of di-jet asymmetry factor $A_J$ for $p+p$ and Pb+Pb collisions at $\sqrt{s_{NN}} = 2.76$~TeV at the LHC. Left panel: $0$-$10$\% centrality; right panel: $10$-$20$\% centrality.}
\label{dijet_Aj}
\end{figure}

In Fig. \ref{dijet_Aj}, we show our calculation of the medium modification of the di-jet asymmetry factor $A_J$ defined as
\begin{equation}
\label{eq:AJ}
A_J = \frac{E_{T,1} - E_{T,2}}{E_{T,1} + E_{T,2}} ,
\end{equation}
where $E_{T,i}, (i=1,2)$ denotes the transverse energy of the leading and sub-leading jet, respectively.
For back-to-back di-jet events in the vacuum, $A_J$ is peaked at zero.
The ATLAS Collaboration measured this quantity by requiring the trigger jet $E_{T,1}>100$~GeV and the second jet in the opposite hemisphere $\Delta\phi>\pi/2$ with $E_{T,2}>25$~GeV.
To proceed, we first generate vacuum di-jet events from PYTHIA \cite{Sjostrand:2007gs} and obtain the distribution for the di-jet asymmetry factor $A_J$ in $p+p$ events.
The modification of each di-jet event in Pb+Pb collisions is obtained as follows.
For each di-jet event, we sample its production points according to the distribution of the binary nucleon-nucleon collisions in collisions of two Pb nuclei.
For asymmetric di-jets ($A_J>0.1$), the trigger bias is taken into account by letting the higher energy jet propagate along the shorter path (implying a smaller energy loss), and the other jet to propagate along the other direction.
For nearly symmetric jet pairs ($A_J<0.1$), such a trigger bias does not apply.

As expected, the number of strongly asymmetric di-jets is significantly increased by the medium evolution which tends to let one jet lose more energy than the other due to the different path lengths of the two jets in the medium.
The asymmetry of di-jets is more prominent in the most central Pb+Pb collisions (left panel of Fig.~\ref{dijet_Aj}) than in mid-central events (right).
The depletion of energy inside the jet cone is a combination of collisional energy loss experienced by all shower partons, radiation outside the jet cone, and the scattering of radiated gluons into angle outside the jet cone.
From our fit to the data we obtain the average path-length weighted transport coefficient in central collisions $\langle \hat{q} \rangle = \langle \hat{q} L \rangle / \langle L \rangle = 1$~GeV$^2$/fm, where the average is over different production points and propagation directions.
This corresponds to a value of $\hat{q}=2.6$~GeV$^2$/fm at the highest temperature $400$~MeV in Au+Au collisions at RHIC,
consistent with the systematic analysis performed in Ref. \cite{Bass:2008rv}.

In summary, we have studied the evolution of a jet shower propagating in a quark-gluon plasma and calculated the loss of energy contained in a given cone angle.
The medium modification of the shower spectrum and shape is described by a differential equation that incorporates both, collisional energy loss and transverse momentum broadening.
Our approach provides a good description of the di-jet asymmetry observed by the ATLAS Collaboration in Pb+Pb collisions at the LHC.
The values of the parton transport coefficients are similar to those describing jet quenching at RHIC, extrapolated to the higher matter density at the LHC.
This suggests that the quark-gluon plasma created at the LHC has similar properties as that studied by the RHIC experiments.

This work was supported in part by Grants No. DE-FG02-05ER41367 and No. DE-SC0005396 from the U.S. Department of Energy.

\bibliographystyle{h-physrev5.bst}
\bibliography{GYQ_refs.bib}

\end{document}